\documentclass[aps,prb,twocolumn,showpacs,superscriptaddress,groupedaddress]{revtex4}

\usepackage{graphicx}
\usepackage{float}
\usepackage{sidecap}
\usepackage{dcolumn}
\usepackage{bm}
\usepackage{amssymb}
\usepackage{amsmath}	
\usepackage{color}
\begin{document}


\widetext

\title{Polarization Response in InAs Quantum Dots: Theoretical Correlation between Composition and Electronic Properties}


\author{Muhammad Usman} \email{usman@alumni.purdue.edu} \affiliation{Tyndall National Institute, Lee Maltings, Dyke Parade, Cork, Ireland.} 
\author{Vittorianna Tasco} \affiliation{National Nanotechnology Laboratory, Istituto Nanoscienze – CNR, Via Arnesano, 73100 Lecce, Italy.}
\author{Maria Teresa Todaro} \affiliation{National Nanotechnology Laboratory, Istituto Nanoscienze – CNR, Via Arnesano, 73100 Lecce, Italy.}
\author{Milena De Giorgi} \affiliation{National Nanotechnology Laboratory, Istituto Nanoscienze – CNR, Via Arnesano, 73100 Lecce, Italy.}
\author{Eoin P. O'Reilly} \affiliation{Tyndall National Institute, Lee Maltings, Dyke Parade, Cork, Ireland.} \affiliation{Department of Physics, University College Cork, Cork Ireland}
\author{Gerhard Klimeck} \affiliation{Purdue University, Network for Computational Nanotechnology, Birck Nanotechnology Center, 1205 W. State Street, West Lafayette, IN 47907, USA.}
\author{Adriana Passaseo} \affiliation{National Nanotechnology Laboratory, Istituto Nanoscienze – CNR, Via Arnesano, 73100 Lecce, Italy.}
\vskip 0.25cm



\begin{abstract}
III-V growth and surface conditions strongly influence the physical structure and resulting optical properties of self-assembled quantum dots (QDs). Beyond the design of a desired active optical wavelength, the polarization response of QDs is of particular interest for optical communications and quantum information science. Previous theoretical studies based on a pure InAs QD model failed to reproduce experimentally observed polarization properties. In this work, multi-million atom simulations are performed to understand the correlation between chemical composition and polarization properties of QDs. A systematic analysis of QD structural parameters leads us to propose a two layer composition model, mimicking In segregation and In-Ga intermixing effects. This model, consistent with mostly accepted compositional findings, allows to accurately fit the experimental PL spectra. The detailed study of QD morphology parameters presented here serves as a tool for using growth dynamics to engineer the strain field inside and around the QD structures, allowing tuning of the polarization response.
\end{abstract}

\maketitle


\textbf{\textit{Introduction:}} Semiconductor nanostructures are being more and more applied to several optoelectronics technologies ranging from lasers\cite{Salhi_1} to optical amplifiers\cite{Akiyama_1} or single photon sources\cite{Dousse_1} where they have successfully overcome critical challenges such as extremely low threshold, high speed response, or entangled photon emission, respectively. In most of these applications a crucial parameter is the polarization response, typically measured in terms of the degree of polarization (DOP = TE-TM/TE+TM)\cite{Usman_1, Usman_2} or the TM/TE ratio\cite{Fortunato_1}. Understanding how this feature is related to the quantum dot (QD) structural symmetry and composition could be very helpful in its tuning. For example, semiconductor optical amplifiers (SOAs) for telecommunications require engineered QDs for isotropic polarization behavior\cite{Akiyama_1, Usman_1, Usman_2}. Polarization control is also crucial for other applications, such as polarization-entangled photons emitted by single QDs\cite{Dousse_1} and polarization sensitive applications of vertical cavity surface emitting lasers\cite{Saito_2, Visimberga_1}.

The structural and electronic properties of III-V QDs epitaxially formed by the self-ordering Stranski-Krastanov (SK) process are strongly affected by surface and growth conditions. Several techniques such as surface structuring\cite{Mohan_1}, nanostructure engineering via strain coupling\cite{Usman_1, Usman_2, Inoue_1, Saito_1, Fortunato_1}, and tuning of growth conditions\cite{Alonso_1} are currently being studied for controlling QD distribution and geometric symmetry. As a common trend, epitaxial InAs nanostructures exhibit asymmetry not only in the growth plane (with respect to [110] and [1$\bar{1}$0] directions)\cite{Usman_1, Usman_2} but also in the vertical plane (along [001])\cite{Villegas_1} due to their preferential flat, lens shape. The vertical confinement leads to a strong compressive biaxial strain suppressing the light-hole (LH) component in the valence-band states leaving mostly the heavy hole (HH) component and resulting in predominantly in-plane (TE) polarized emission from the interband transitions\cite{Koudinov_1}.

\begin{figure*}
\centering
\includegraphics[width=1.0\textwidth]
{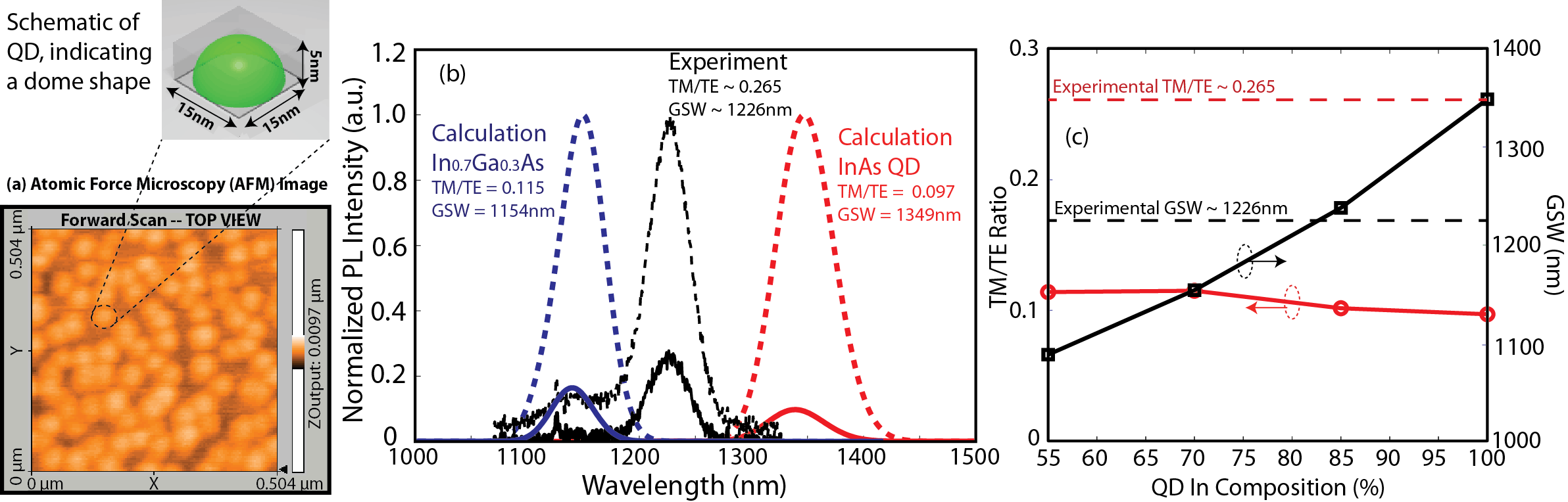}
\caption{(a) Atomic force microscopy (AFM) image of a single QD layer sample before GaAs capping revealing a dot density of 3.5x10$^{10}$ dots/cm$^{2}$ with average height of 5nm as indicated in the inset above the figure. (b) Normalized PL intensity is plotted for TE$_{100}$ (broken lines) and TM$_{001}$ modes (solid lines), from experimental measurements and theoretical calculations as a function of wavelength. The calculations are done for a pure InAs and an In$_{0.7}$Ga$_{0.3}$As random alloy quantum dot. The comparison of the experimental measurement and the theoretical calculations clearly indicates a significant difference for the TM/TE ratios. (c) Plots of the ground state wavelengths (GSW) and TM/TE ratio as a function of the In-composition of the QD. As the In composition decreases from 100$\%$ (pure InAs QD) to 55$\%$ (an alloyed In$_{0.55}$Ga$_{0.45}$As QD), the GSW decreases drastically whereas the TM/TE ratio only slowly increases. The experimental values of the GSW and TM/TE ratio are marked as dashed lines. Clearly a single In-composition in the uniform composition model is unable to simultaneously reproduce experimental values of the GSW and TM/TE ratio.}
\label{fig:Fig1}
\end{figure*}
\vspace{2mm}

InAs QDs obtained via the SK process have been found to be significantly influenced by In-Ga intermixing and In-segregation effects during the capping and post-growth annealing processes\cite{Medhekar_1, Shumway_1, Fry_1, Barker_1}. Several composition profiles have been proposed and different investigation techniques employing high resolution transmission electron microscopy, x-ray diffraction, photoelectron microscopy or scanning probe microscopy have been used to exactly map InAs nanostructures\cite{Kegel_1, Liu_1, Lemaitre_1}. As recently reviewed by Biasiol and Heun\cite{Biasoil_1}, results in the literature do not lead to an unified model, where the actual composition profile of the QDs is strongly related to the growth conditions. Their review highlights a common tendency that the chemical composition of a typical SK QD has gradients both along the growth and the in-plane directions: the In composition increasing from base to top due to In segregation effects and decreasing from the center towards the edges in the lateral directions due to In-Ga intermixing effects. However, such a complex structure has not been considered so far to theoretically understand the polarization response of InAs QDs. 

Previous theoretical\cite{Usman_1, Saito_1, Sheng_1} studies of QD polarization response are based on a pure InAs type QD composition profile, thus significantly limiting their accuracy and leading to discrepancy between theory and experiment. A clear example of such a discrepancy is the failure to reproduce the large values of the experimentally measured DOP by using both $\textbf{k} \cdot \textbf{p}$\cite{Saito_1, Sheng_1} and atomistic tight binding methods\cite{Usman_1}. Thus a better understanding of the correlation between the QD structure and the degree of polarization remains an outstanding challenge.

In this letter, the polarization response of InAs QDs is theoretically studied by atomistic simulations, introducing a compositional model capable of fitting experimental measurements both of the electronic transitions and of the TM/TE ratio. The actual complex composition and geometry of SK InAs QDs is mimicked by a two composition QD model, reproducing the experimentally measured polarization behavior of a single QD layer and highlighting the relevance of atomic scale processes like segregation and intermixing.

\begin{figure*}
\centering
\includegraphics[width=1.0\textwidth]
{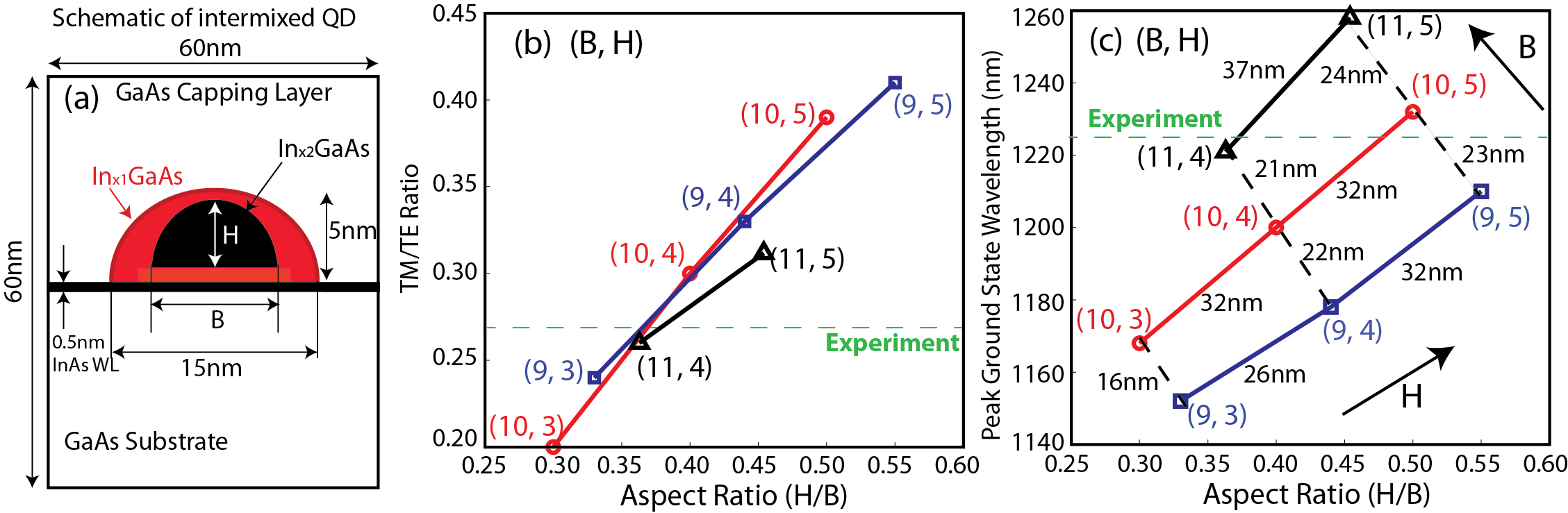}
\caption{(a)Schematic diagram of the two composition QD model, placed on top of a 0.5nm thick InAs wetting layer (WL). The QD has overall dimensions of base=15nm and height=5nm in accordance with the TEM\cite{Raino_1}. It consists of two regions. The central In-rich region with In$_{x2}$Ga$_{1-x2}$As composition (x2 $\geq$ 80$\%$) is of size base=B and height=H. The outer In-poor region is made up of In$_{x1}$Ga$_{1-x1}$As (x1 $\leq$ 40$\%$) material. (b) The TM/TE ratio as a function of the aspect ratio (AR=H/B) for various values, (B, H), in units of nm. The compositions x1 and x2 are 40$\%$ and 100$\%$, respectively. (c) The GSW as a function of the aspect ratio (AR=H/B) for various values of H and B. The compositions x1 and x2 are 40$\%$ and 100$\%$, respectively. The numbers in nm along the solid and dotted lines indicate the change in the value of GSW between the adjacent points along each line.}
\label{fig:Fig2}
\end{figure*}
\vspace{2mm}  

\textbf{\textit{Experimental procedure:}} The QD samples used in this study were grown on semi-insulating GaAs substrates, by a COMPACT 21- Riber Molecular Beam Epitaxy (MBE) system equipped with a reflection high energy electron diffraction (RHEED) gun to monitor the surface evolution in-situ during growth. After growth of a GaAs buffer layer at 600$^\circ$C, the substrate temperature was lowered down to 500$^\circ$C and QDs were formed by covering the buffer with 2.8 MLs of InAs. The 2D-3D growth mode transition is demonstrated by the RHEED pattern evolving from streaky-like to spot-like after deposition of 1.7 MLs of InAs. Afterwards, dots were immediately capped by a GaAs spacer layer grown at the same low temperature. The single layer QD sample was then covered with a 20 nm GaAs cap terminating the structure. An uncapped single QD layer sample was also grown under the same conditions and its morphology analysed by atomic force microscopy (AFM), providing a dot density of 3.5x10$^{10}$ dots/cm$^{2}$ with average height of 5nm, as shown in Figure \ref{fig:Fig1}(a). The GaAs capped QDs were estimated by transmission electron microscopy (TEM) to have a dome-like shape with base diameter $\approx$ 15nm and height $\approx$ 5nm (see inset above Fig. ~\ref{fig:Fig1}(a)).

For investigating the polarization behavior, samples were excited from the top with a cw Ar+ laser ($\lambda$ = 514 nm). The room temperature photoluminescence (PL) signal from the cleaved edge of the samples was first collected by a long focal lens (200 mm) and it was then filtered by a linear polarizer and focused by a second lens into the monochromator. 
 
\textbf{\textit{Theoretical model:}} The theoretical modeling is performed using NEMO 3-D\cite{Klimeck_1, Klimeck_2}. NEMO 3-D is an atomistic simulator based on the valence force field (VFF) method\cite{Keating_1} for strain calculations and the twenty band $sp^3d^5s^*$ tight binding model\cite{Boykin_1} for the electronic structure. NEMO 3-D is a multi-scale simulator capable of performing multi-million atom simulations for realistic QD dimensions surrounded by large GaAs buffers to model the long range impact of the strain and piezoelectric potential. This tool has already been used to model QD structures providing results in good agreement with experimental data\cite{Usman_1, Usman_2, Usman_3, Usman_4, Lee_1}. Piezoelectric potentials, both linear and quadratic, are calculated by solving Poison's equation according to a published recipe\cite{Usman_4} and are included in the calculations of the electronic spectra. The inter-band optical transition strengths are calculated using Fermi's golden rule, with the squared magnitude of the optical matrix elements found by summing over the spin degenerate states. The polarization dependent TE and TM spectra are calculated along the [100] and [001] directions respectively, as a cumulative sum of optical transitions between the lowest conduction band energy level (E1) and the highest four valence band energy levels (H1, H2, H3, and H4), where each transition strength is artificially broadened by multiplication with a Gaussian distribution centered at the wavelength of the transition\cite{Usman_1}. The highest four hole energy levels are chosen here instead of only the top most level (H1), because the valence band states are closely spaced on the energy scale (H1-H4 $\leq$ kT $\approx$ 26meV at T=300K) and multiple hole states contribute to the ground state optical emission peak at the room temperature\cite{Usman_2}.

\textbf{\textit{Model 1: uniform compositions:}} We start our analysis by first assuming a uniform material profile throughout the QD geometry as in previous theoretical studies of the polarization response\cite{Usman_1, Saito_1, Usman_2, Sheng_1, Sheng_2, Williamson_1}. Dots are assumed to be lens-shaped with base diameter and height set at 15nm and 5nm, respectively, in accordance with the experimental TEM analysis\cite{Raino_1}. Figure \ref{fig:Fig1}(b) plots the experimental normalized PL spectra taken under the two polarization conditions (TE and TM) and compares them with the theoretically calculated spectra for both the ideal case of a pure InAs QD and an In$_{0.7}$Ga$_{0.3}$As random alloy QD. The InGaAs alloy configuration has been suggested by some previous theoretical studies\cite{Sheng_1, Sheng_2, Williamson_1} to mimic the In-Ga intermixing effect.

\begin{SCfigure*}
\centering
\includegraphics[width=0.65\textwidth]%
{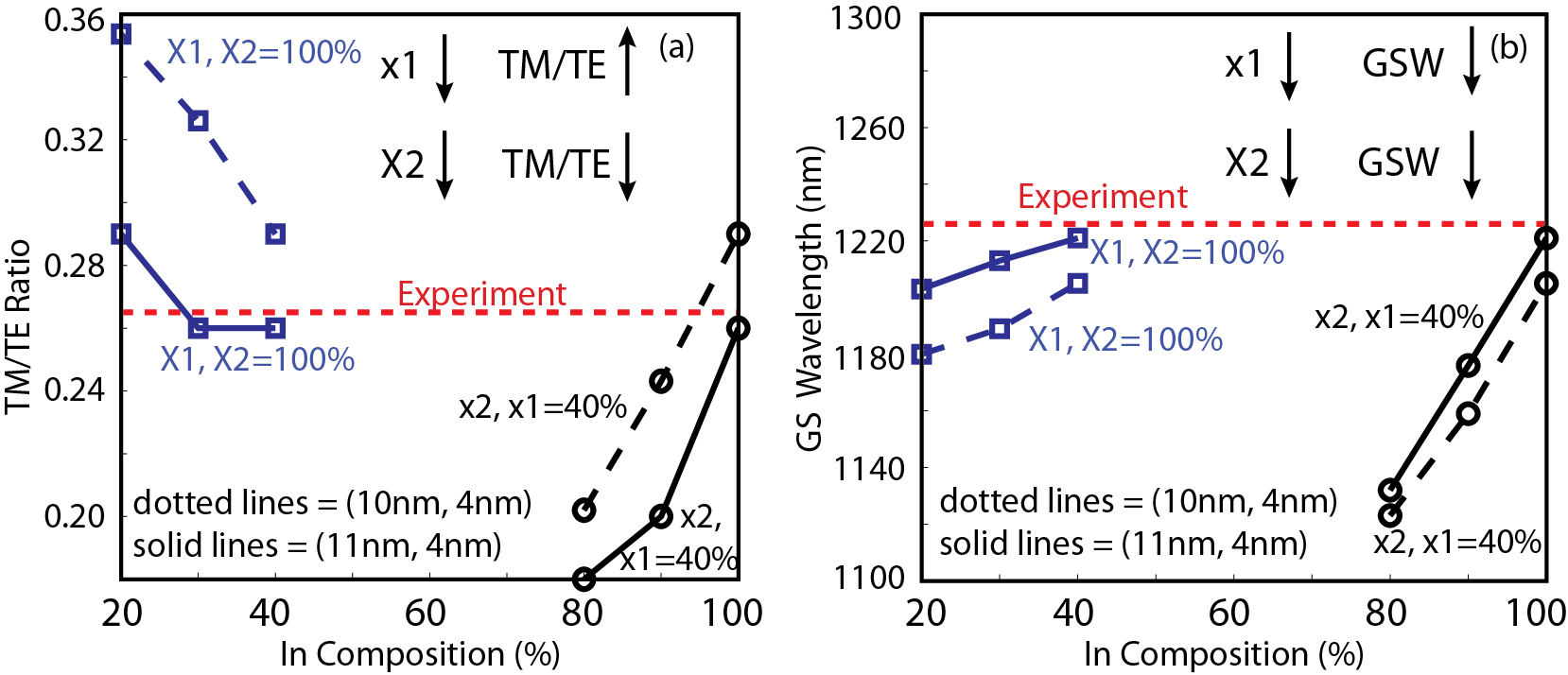}
\caption{(a)Plots of the TM/TE ratio as a function of the compositions x1 and x2 as described in the QD schematic of Fig. ~\ref{fig:Fig2}(a). The plots are for two dimensions of the core region: (B, H) = (10nm, 4nm) and (11nm, 4nm). The arrows indicate an increase/decrease of the TM/TE ratio with respect to a decrease in the compositions x1 and x2. (b) Plots of the GSW as a function of the compositions x1 and x2. The arrows indicate increase/decrease of the GSW with respect to a decrease in the compositions x1 and x2.}
\label{fig:Fig3}
\end{SCfigure*}
\vspace{2mm} 

The pure InAs QD model clearly fails to reproduce the experimentally measured spectra. The TM/TE ratio is calculated to be 0.097 which is significantly lower than the experimentally measured value of $\approx$0.265. The peak wavelength in the PL spectra, hereafter referred to as ground state wavelength (GSW), is also overestimated by $\approx$123nm. This is consistent with the earlier theoretical studies\cite{Usman_1, Saito_1, Sheng_1} where a pure-InAs-QD type model predicted TE-dominant optical emissions and failed to match the experimentally measured PL spectra. Lowering the average In composition inside the QD to 70$\%$ to mimic the effect of In-Ga intermixing during the capping process increases the TM/TE ratio to 0.115 due to the reduced biaxial strain, but at the same time blue shifts the GSW by 72nm with respect to the experimental value of 1226nm. To further investigate the impact of lowering the In composition of the QD, we gradually reduce its value from 100$\%$ (pure InAs QD) to 55$\%$ (In$_{0.55}$Ga$_{0.45}$As QD) and plot the corresponding values of the GSW and TM/TE ratio in Figure ~\ref{fig:Fig1}(c). As the In composition reduces, the GSW decreases to 1088nm, crossing the experimental value (shown by the dotted line) for In $\approx$84$\%$. However, the TM/TE ratio increases very slowly, only reaching a value of 0.114 for In=55$\%$. This clearly indicates that an uniform In-composition, either as pure InAs or some InGaAs alloy composition, cannot be selected as the chemical composition of the QD when seeking agreement between theory and experiment for both the GSW and TM/TE ratio. Therefore, a more complex compositional model is required in order to understand the experimental results.   

Several experimental\cite{Biasoil_1, Lemaitre_1, Passow_1, He_1, Chung_1, Goldman_1, Lita_1} and theoretical\cite{Shumway_1, Medhekar_1, Migliorato_1, Migliorato_2, Yang_1} studies have shown that In-Ga intermixing and In-segregation effects significantly influence the chemical composition of QDs during the growth of the GaAs capping layer. Thus we hypothesize that a quantitative agreement between theory and experiment can be achieved if some inter-diffusion of Ga atoms inside the QD region is included in the model. The most widely suggested profile\cite{Biasoil_1, Lemaitre_1, Passow_1, He_1, Chung_1} consists of an In-rich region along the vertical direction and a decreasing In-composition laterally from the center to the border. Such a complex configuration can be theoretically approximated by defining two composition regions inside the QD: an In-rich core surrounded by a low In-concentration region. Following these considerations, we now analyse the polarization response of such a QD model and explore the structural parameters to reproduce the experimental measurements. 

\textbf{\textit{Model 2: non-uniform composition:}} In order to get the most reliable configuration, as close as possible to the experimental one\cite{Biasoil_1} a simplified double region scheme is adopted, consisting of an In-rich inner core of  In$_{x2}$Ga$_{1-x2}$As (x2 typically $\geq$ 80$\%$) material, surrounded by an In-poor thin region of In$_{x1}$Ga$_{1-x1}$As (x1 typically $\leq$40$\%$), as schematically shown in Figure 2(a). These ranges for the compositions x1 and x2 have been chosen from previous experimental results\cite{Lemaitre_1, Passow_1}. The actual size of the In-rich core, defined by inner region base width (B) and height (H) as shown in Figure ~\ref{fig:Fig2}(a), strongly depends on the growth dynamics of the capping layer. The overall size of QD obtained from TEM\cite{Raino_1} places constraints on the values of B and H: H $\leq$ 5nm and B $\leq$ 15nm. On the basis of this QD model, we perform a systematic investigation by changing the parameters B, H, x1, and x2 and comparing the corresponding GSW, TE, and TM modes with the experimental measurements. Finally, it is worth noting that the two cases studied earlier in the uniform composition model, an InAs QD and an In$_{0.7}$Ga$_{0.3}$As QD, can be included by this two composition model by assuming x1=x2=100$\%$ and x1=x2=70$\%$, respectively.  

\textbf{\textit{Variations of B and H:}} First, in order to investigate the impact of the parameters H and B on the polarization response, we choose x2=100$\%$ and x1=40$\%$. Such an assumption is motivated by considering that the central core of InAs nano-islands is always reported as being In-rich and that the investigations reported so far indicate that the maximum value for the In-composition can reach 100$\%$\cite{Biasoil_1}. 

Figures \ref{fig:Fig2}(b) and \ref{fig:Fig2}(c) plot the TM/TE ratio and the GSW, respectively, as a function of the aspect ratio (AR = H/B) of the inner core. The target experimental values are shown as horizontal dotted lines. The TM/TE ratio in Figure ~\ref{fig:Fig2}(b) linearly increases with AR when H is increased for a fixed value of B. This is because an increase in H will result in a reduction of the biaxial strain close to the center of the QD which will reduce the splitting between the HH and LH bands. As a result, the magnitude of the TM-mode will increase, thus increasing the TM/TE ratio. However, our calculations show that the slope of the TM/TE ratio as a function of AR=H/B is different for different values of B. In general, it decreases as B increases, indicating a decreasing impact of H on the TM/TE ratio. This is consistent with a previous theoretical study\cite{Schliwa_1} where hole energy levels in smaller QDs were found to be more sensitive to changes in QD AR as compared to larger QDs. We want to highlight here that our theoretical calculations show that even changes of 1-2nm in B or H values, which can actually be induced by highly controlled growth dynamics, result in a drastic change in the polarization properties of the QD sample.  

The plots of GSW as a function of AR in Figure ~\ref{fig:Fig2}(c) exhibit a linear dependence of GSW on H for a fixed value of the B (shown by the solid lines), as well as a linear dependence of the GSW on the value of B for a fixed value of H (shown by the dotted lines). The analysis of Figure ~\ref{fig:Fig2}(b) and (c) indicates that two sets of values for (B, H), namely (10nm, 4nm) and (11nm, 4nm), give transition wavelengths and a TM/TE ratio close to the experimental values. A slight change in the sets of values of B and H ((10nm, 5nm) and (9nm, 5nm)) leads to a GSW value close to the experimental one, but introduces a discrepancy in the corresponding TM/TE ratios. We therefore choose (B, H) to be (10nm, 4nm) and (11nm, 4nm) to analyse in the following sections the impact that varying x1 and x2 has on the calculated values of the GSW and TM/TE ratio.

\textbf{\textit{Variation in x1:}} The outer shell composition x1 was assumed above to be 40$\%$ in the study of the effect of the B and H parameters on the QD optical properties. However, this composition can have a lower value depending upon the growth dynamics during the growth of the GaAs capping layer. We therefore decrease this composition to 30$\%$ and 20$\%$ for both (10nm, 4nm) and (11nm, 4nm) dimensions, while keeping x2 at 100$\%$. Figures ~\ref{fig:Fig3}(a) and ~\ref{fig:Fig3}(b) show the impact of such variations in x1 on the TM/TE ratio and GSW, respectively. It is clear that decreasing the value of x1 has opposite impact on the TM/TE ratio and GSW: the value of TM/TE ratio increases and the value of the GSW decreases. Following these trends for the dimensions (B, H) plotted in Figures ~\ref{fig:Fig2}(b) and ~\ref{fig:Fig2}(c), we deduce that a decreasing GSW with respect to x1 will only bring (11nm, 5nm) and (10nm, 5nm) closer to the experimental GSW. However, the TM/TE ratio for these two dimensions is already higher than the experimental value (see Figure ~\ref{fig:Fig2}(b)), so a corresponding increase in the TM/TE ratio will further reduce their agreement with the experimental TM/TE ratio. Therefore, we deduce that decreasing x1 below 40$\%$ reduces agreement with the experimental values (either GSW or TM/TE ratio) for all of the dimensions (B, H) and hence we require the poor In-composition region comprising QD outer shell to have a composition close to 40$\%$. 

\textbf{\textit{Variation in x2:}} Next, we analyse the impact of the last unknown parameter x2 in our proposed two compositional model (see Figure ~\ref{fig:Fig2}(a)), keeping x1=40$\%$. Figures ~\ref{fig:Fig3}(b) and ~\ref{fig:Fig3}(c) plots variations in the TM/TE ratio and GSW, respectively, when x2 is decreased from 100$\%$ to 90$\%$ and 80$\%$. Contrary to x1, a decrease in x2 has similar impact on both the TM/TE ratio and GSW: both decrease as the composition x2 decreases. From Figures ~\ref{fig:Fig2}(b, c), these trends imply that the values of both the TM/TE ratio and GSW will become closer to the experimental values only for the two dimensions, (11nm, 5nm) and (10nm, 5nm). However, the GSW for the dimensions (10nm, 5nm) is very close to the experimental GSW whereas the TM/TE ratio for this dimension is significantly different from the experimental value. Hence a collective agreement of both the GSW and TM/TE ratio with the experimental values is not possible for a decrease in x2. For example, at x2=90$\%$, the GSW decreases to 1187nm ($\approx$34nm below the experiment value) whereas the TM/TE ratio is still 0.31 ($\approx$0.045 above the experimental value). For the (B, H) = (11nm, 5nm) case, if x2 decreases to 90$\%$, the GSW and TM/TE ratio decrease to 1208nm and 0.27, respectively, which are in reasonable agreement with the corresponding experimental values of 1221nm and 0.265. 

We complete our analysis of QD compositions by considering simultaneous decrease in x1 and x2 and finding that their cumulative impact on the GSW and TM/TE values is roughly equal to the sum of their individual effects. We find that the TM/TE ratio is more sensitive to x2 as compared to x1, and therefore any simultaneous decrease in the compositions x1 and x2 will reduce the values of both parameters (GSW and TM/TE ratio).

\begin{figure}
\centering
\includegraphics[width=0.46\textwidth]%
{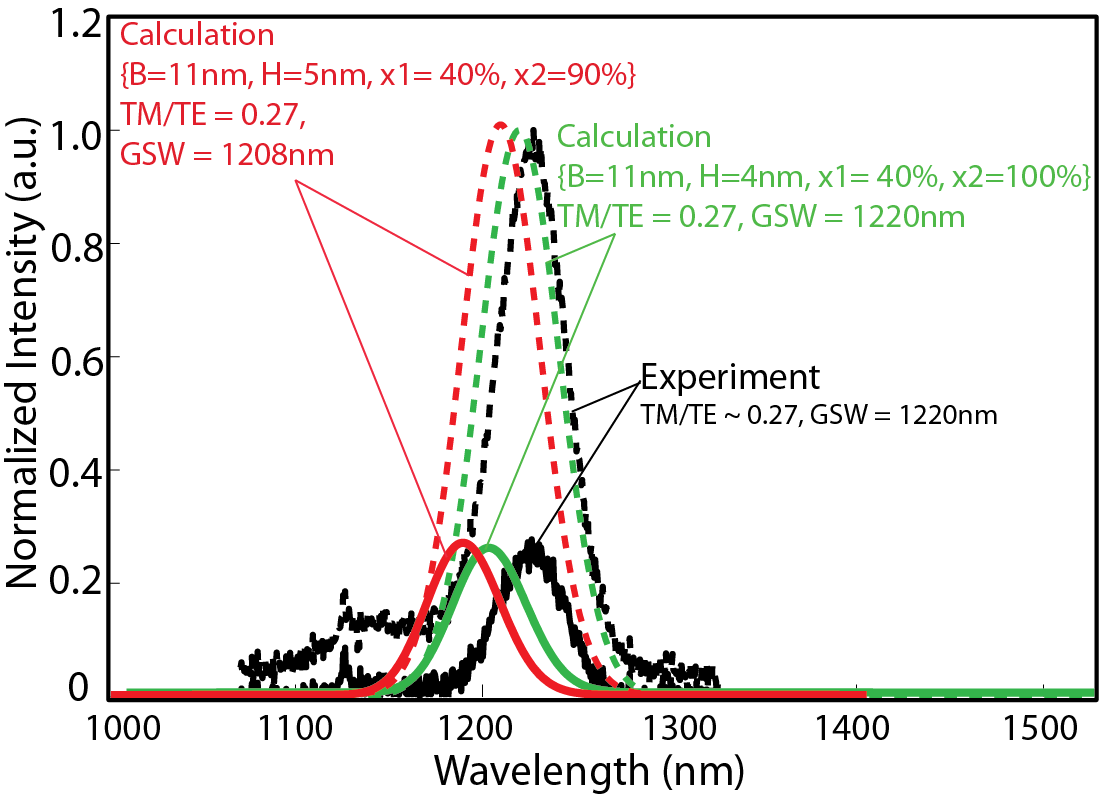}
\caption{Normalized PL spectra is shown from the experimental measurement (black lines) and compared with the theoretical calculations (green and red lines). Both TE (dotted lines) and TM (solid lines) components are plotted. A slight difference between the GSW of the calculated TM and TE modes is due to the fact that the TE mode is dominantly from E1-H1 transition and the TM mode is dominantly from E1-H3 and E1-H4 transitions. Such a difference will be difficult to observe in the experiment, because of the broadening of the PL spectra which are typically collected from multiple QDs.}
\label{fig:Fig4}
\end{figure}
\vspace{2mm}

\textbf{\textit{QD structural parameters:}} From the preceding discussions on the effect of B, H, x1, and x2 in the QD structure of the Figure ~\ref{fig:Fig2}(a), we conclude that for our experimentally measured QD sample, the most effective model is represented by the following parameters: B $\approx$ 11nm, H $\approx$ 4-5nm, x1 $\approx$ 40$\%$, and x2 $\approx$ 90-100 $\%$. The calculated PL spectra for based on these sets of parameters are compared with the experimental PL spectra in Figure ~\ref{fig:Fig4}. It is worth noting that the proposed structures roughly reproduce the complex composition profile experimentally found by many authors\cite{Biasoil_1}, including the anisotropic In-Ga intermixing behavior with respect to the crystallographic direction, with a larger in-plane inter-diffusion region (nearly 2nm around the base of the dot) and a negligible inter-diffusion at the top (0-1nm) where segregation predominated. 

\begin{figure}
\centering
\includegraphics[width=0.5\textwidth]%
{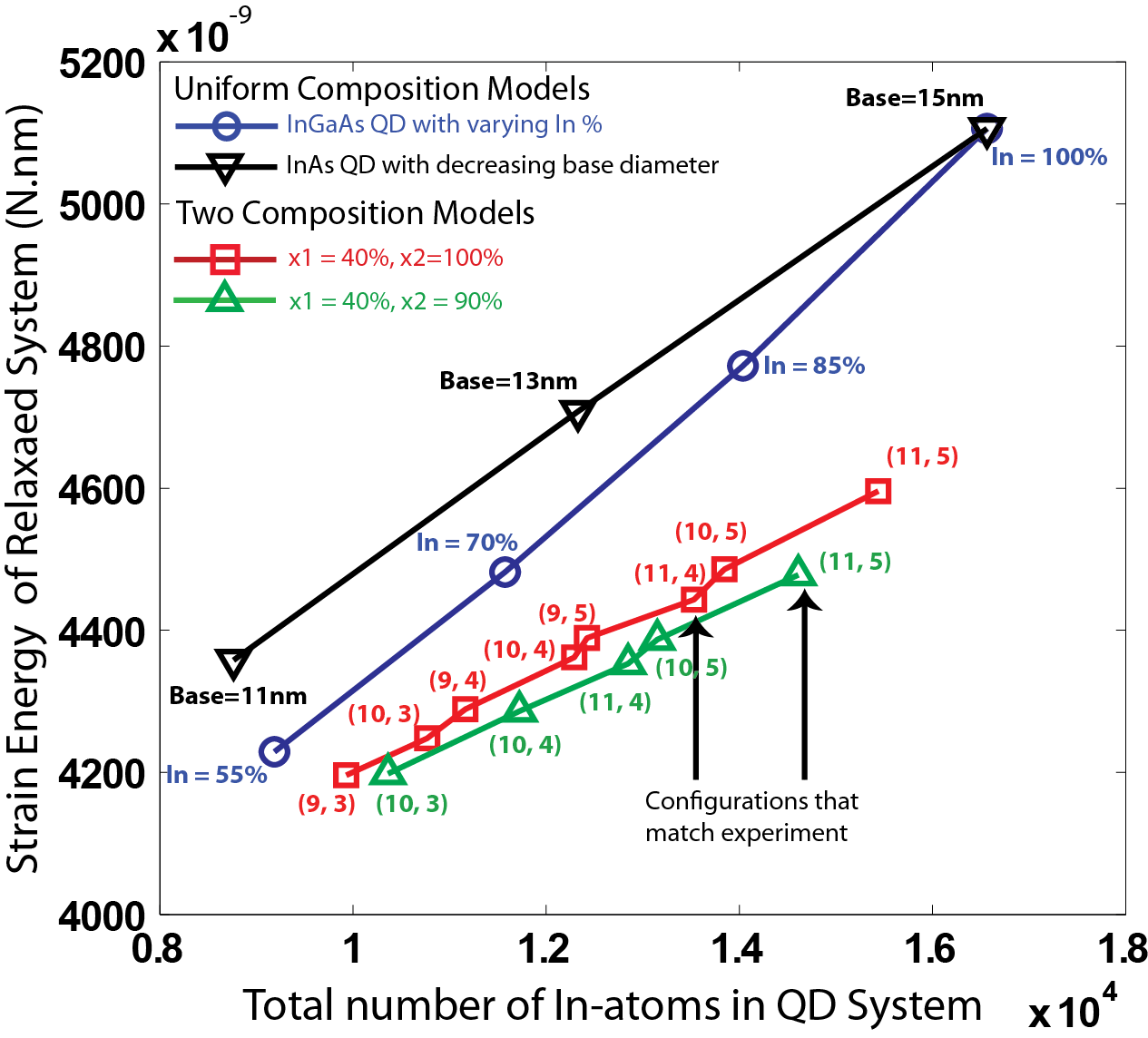}
\caption{Plots of the total strain energy of the relaxed QD systems as a function of the total number of In atoms in the system. Both the uniform and non-uniform composition models have been considered. The relaxed strain energies for the non-uniform composition configurations are smaller than the relaxed strain energies of the uniform composition configurations with similar number of In atoms.}
\label{fig:Fig6}
\end{figure}
\vspace{2mm}  

\textbf{\textit{Relaxed strain energies:}} The QD systems have been relaxed using the VFF model to reach a minimum strain energy configuration\cite{Klimeck_1, Keating_1}. The comparison of the relaxed strain energies for the various QD systems under study may provide additional insight to find a more likely experimental geometry. Figure ~\ref{fig:Fig6} plots the relaxed strain energies as a function of the total number of In-atoms for the various configurations after the VFF minimization is achieved. We consider two cases of the uniform composition model: an In$_{x}$Ga$_{1-x}$As QD with decreasing In-composition (see Figure ~\ref{fig:Fig1}(c)) and a pure InAs QD with decreasing base diameter. We also consider two cases of the non-uniform composition model: (B, H, x1=40$\%$, x2=100$\%$) and (B, H, x1=40$\%$, x2=90$\%$). Clearly, the relaxed strain energies belonging to the two layer composition configurations are lower than the strain energies of the uniform composition configurations for a similar number of In atoms. This suggests that a two composition model is more favourable as a proposed experimental geometry when compared to a uniform composition model, in accordance with our earlier findings.      

\begin{figure}
\includegraphics[scale=0.32]{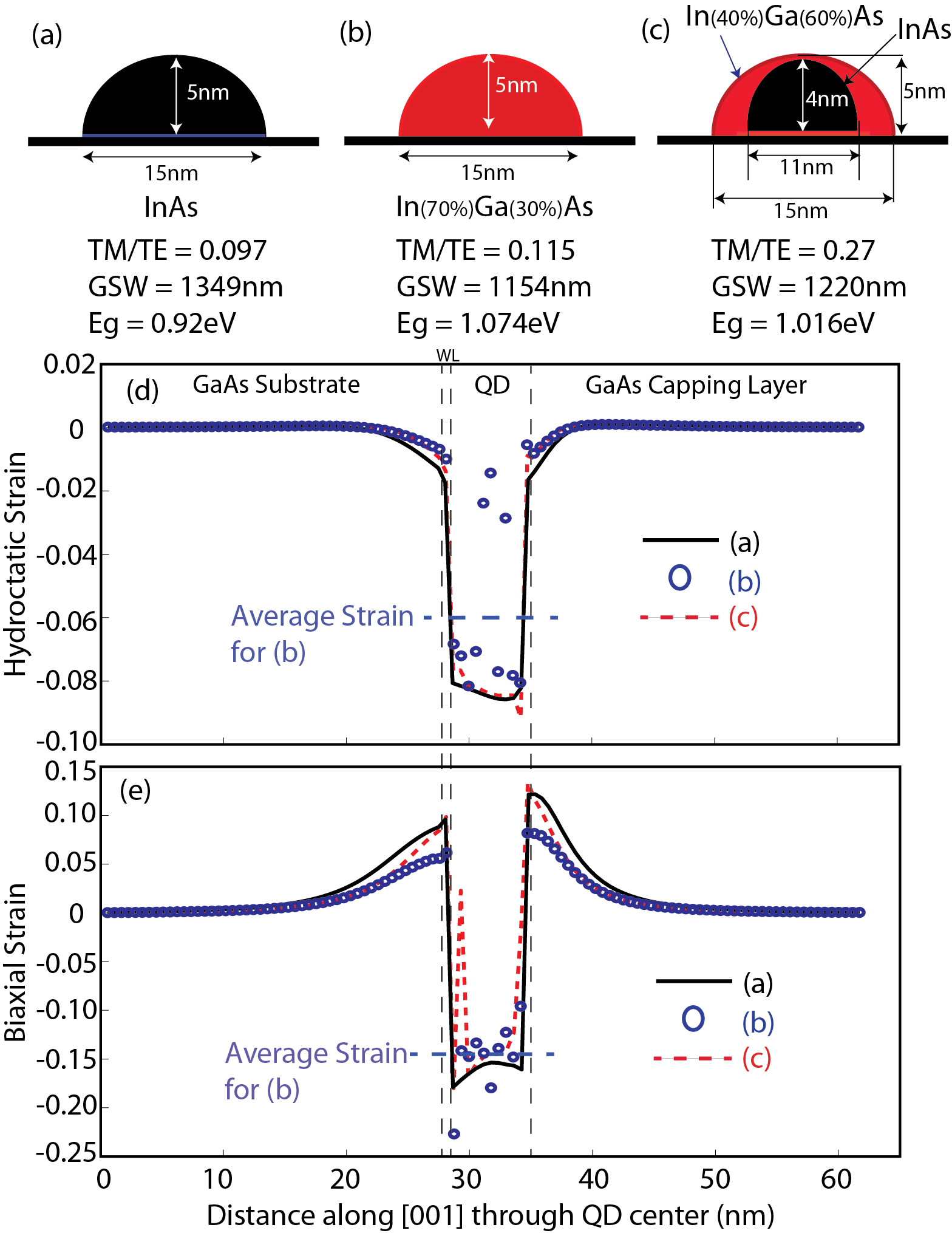}
\caption{(a-c) Schematics of three QD models under study: (a) A pure InAs QD, (b) A random alloy In$_{0.7}$Ga$_{0.3}$As QD, and (c) A QD with InAs core surrounded by an In$_{0.4}$Ga$_{0.6}$As region. (d, e) Plots of the Hydrostatic ($\epsilon_{H} = \epsilon_{xx}+\epsilon_{yy}+\epsilon_{zz}$) and Biaxial ($\epsilon_{B} = \epsilon_{xx}+\epsilon_{yy}-2\epsilon_{zz}$) strain components along the [001] direction through the center of the QDs. The solid lines, open circles, and dotted lines are for the QD models shown in (a), (b) and (c), respectively. The horizontal dotted lines are marked to indicate the average values of the strain inside the QD region for the In$_{0.7}$Ga$_{0.3}$As alloy QD model.}
\label{fig:Fig5}
\end{figure}

\textbf{\textit{Strain profile analysis:}} Our model 2 agrees with the experimental TE and TM PL spectra. However a comparison of strain profiles is required to further understand the difference between the calculated PL spectra for models 1 and 2. In Figures ~\ref{fig:Fig5}(a-c), we compare three QD configurations: (i) a pure InAs QD with base=15nm and height=5nm, (ii) an In$_{0.7}$Ga$_{0.3}$As alloy QD of the same size, and (iii) a two-layer composition as in Figure \ref{fig:Fig2}(a) with B=11nm, H=4nm, x1=40$\%$, and x2=100$\%$ which gives the best agreement with experiment. For reference, the previously calculated values of the GSW and TM/TE ratio are also listed underneath. To understand the shifts in the values of the TM/TE ratio and GSW in (b) and (c) with respect to (a), we compare the hydrostatic and biaxial strain plots along the [001] direction through the center of the QDs in Figures ~\ref{fig:Fig5}(d) and (e), respectively. For the In$_{0.7}$Ga$_{0.3}$As alloy QD, the strain values inside the QD region are randomly distributed due to the disordered composition of the alloy. We mark the average values of the hydrostatic and biaxial strains by the dotted horizontal lines. 

The hydrostatic strain shifts the conduction and valence band edges, therefore increasing the band gap, whereas the biaxial strain is mainly responsible for controlling the splitting between the HH and LH valence band edges\cite{Usman_3}. The strain driven band edge shifts in Table ~\ref{tab:table1} are calculated from simple analytical expressions involving InAs/GaAs deformation potentials\cite{Usman_3}.  

\vspace{1mm}    

\textbf{\textit{Blue shift of the GSW:}} The blue shifts in the GSW for (b) and (c) with respect to (a) indicate increase of the optical gaps (E$_{g}$), which can be understood as a cumulative affect of the relaxed hydrostatic strains and the increase of the band gaps due to the reduced average In compositions of the QDs~\cite{Williamson_1}. By comparing the hydrostatic strain profiles in Figure ~\ref{fig:Fig5} (d), we deduce that the presence of InGaAs alloy, in general, relaxes the strain which can be attributed to reduced QD lattice mismatch with the GaAs buffer. As a result, the band gaps for (b) and (c) will be smaller when compared to (a) due to smaller strain driven shifts in the conduction and heavy hole band edges. For example, from Table ~\ref{tab:table1}, the strain driven shifts predict that the band gap for the In$_{0.7}$Ga$_{0.3}$As QD is 373.7-231.6=142.1meV smaller than the InAs QD. However if the relative increase ($\approx$310meV) in the band gap of the In$_{0.7}$Ga$_{0.3}$As QD due to the reduced average In composition is also added, the cumulative change is 310-142.1=167.9meV increase with respect to the InAs QD, consistent with the increase of the optical gap. To summarize, the relaxation of the hydrostatic strain due to presence of the InGaAs alloy in (b) and (c) introduces reductions in the band gaps which are smaller than the corresponding band gap increase due to the reduced average In compositions, and hence the cumulative impact of these two affects results in overall increase of the optical/band gaps (blue shifts of the GSWs) for (b) and (c) with respect to (a). 

\begin{table}
\caption{\label{tab:table1}The impact of the hydrostatic ($\epsilon_{H}$) and biaxial ($\epsilon_{B}$) strains on the conduction (E$_{c}$), heavy hole (E$_{HH}$), and light hole (E$_{LH}$) band edges are given for the QD models shown in Figures \ref{fig:Fig5}(a, b, and c). The band edge deformations ($\delta$E$_{CB}$, $\delta$E$_{HH}$, and $\delta$E$_{LH}$) are calculated from the analytical expressions given in the Ref. 29. $\delta$E$_{gs}$=$\delta$E$_{CB-HH}$ and $\delta$E$_{H}$=$\delta$E$_{HH-LH}$ represent the changes in the band gap and the HH/LH splittings due to the strain.}
\begin{ruledtabular}
\begin{tabular}{cccccccc}
QD & $\epsilon_{H}$ & $\epsilon_{B}$ & $\delta$E$_{CB}$ & $\delta$E$_{HH}$ & $\delta$E$_{LH}$ & $\delta$E$_{gs}$ & $\delta$E$_{H}$ \\
\cline{2-8}
Model & &  &  meV & meV & meV & meV & meV \\
 \hline \\
(a) & -0.0842 & -0.1536 & 427.7 & 54 & -222.4 & 373.7 & 278.4 \\
(b) & -0.06 & -0.148 & 304.8 & 73.2 & -193.2 & 231.6 & 266.4 \\
(c) & -0.0818 & -0.1449 & 415.5 & 48.6 & -212.2 & 366.9 & 260.8 \\
\end{tabular}
\end{ruledtabular}
\end{table}

\textbf{\textit{Biaxial strain relaxation increases TM/TE Ratio:}} Figure ~\ref{fig:Fig5}(e) compares the biaxial strain components and Table ~\ref{tab:table1} provides the values of the corresponding changes in the HH and LH band edges for the three QD configurations. When compared to the pure InAs QD, the biaxial strain component is reduced for both the InGaAs alloy QD and the 2-composition QD model. This reduces the splitting between the HH and LH bands, thus increasing the LH component in the top most valence band states. A past theoretical investigation\cite{Narvaez_1} has also reported an increased HH/LH intermixing for an InGaAs QD when compared to a pure InAs QD. The increased LH character in turn enhances the TM component\cite{Usman_1}, as we compute for configurations (b) and (c). The smallest HH/LH splitting responsible for the enhanced TM component occurs for the two-layer composition model. 
                    

\textbf{\textit{Conclusions:}} In conclusion, multi-million atom simulations are performed to understand the correlation between the chemical composition of self-assembled QDs and their electronic properties by reproducing the experimentally measured polarization properties. A single composition QD model, being either a pure InAs QD or an alloyed InGaAs QD, failed to reproduce the experiment. To fully understand the experimental polarization behavior, the model must take In-Ga intermixing effects into account, by representing the QDs with variable compositions. Based on a systematic analysis of the dependence of the TM/TE ratio on the various QD morphology parameters, we propose a two composition model that accurately reproduces the measured PL response. The model gives results consistent with the experimental compositional findings and highlights the strong anisotropy of atomic scale phenomena like intermixing and segregation affecting the polarization behavior of these nanostructures. These results could represent a tool for using growth dynamics to engineer the strain field inside and around the QD structures, allowing tuning of the polarization properties, a critical parameter for several challenging applications.


\textbf{\textit{Acknowledgements:}} Computational resources from National Science Foundation (NSF) funded Network for Computational Nanotechnology (NCN) through \url{http://nanohub.org} and Rosen Center for Advanced Computing(RCAC), Purdue University are acknowledged. MU and EPOR acknowledge financial support from European Union project BIANCHO (FP7-257974) and from Science Foundation Ireland (SFI). 


\end{document}